\documentclass[aps,pre,showpacs]{revtex4}

\usepackage{amssymb,amsmath}
\usepackage[pdftex]{graphicx}

\newcommand{\bd}[1]{\mbox{\boldmath$#1$}}

\newcommand{\Pd}[1]{\ensuremath{\Bigl[\!\!\Bigl[#1\Bigr]\!\!\Bigr]}}

\begin{document}

\title{Noise-induced synchronization of oscillatory convection and its optimization}

\author{Yoji Kawamura}
\email{ykawamura@jamstec.go.jp}
\affiliation{Institute for Research on Earth Evolution,
Japan Agency for Marine-Earth Science and Technology, Yokohama 236-0001, Japan}

\author{Hiroya Nakao}
\affiliation{Department of Mechanical and Environmental Informatics, Tokyo Institute of Technology, Tokyo 152-8552, Japan}

\date{January 17, 2014} 

\pacs{05.45.Xt, 05.40.Ca, 82.40.Bj, 82.40.Ck}

\begin{abstract}
  We investigate common-noise-induced phase synchronization
  between uncoupled identical Hele-Shaw cells exhibiting oscillatory convection.
  Using the phase description method for oscillatory convection,
  we demonstrate that the uncoupled systems of oscillatory Hele-Shaw convection
  can exhibit in-phase synchronization when driven by weak common noise.
  We derive the Lyapunov exponent determining the relaxation time for the synchronization,
  and develop a method for obtaining the optimal spatial pattern of the common noise
  to achieve synchronization.
  The theoretical results are confirmed by direct numerical simulations.
\end{abstract}

\maketitle


\section{Introduction} \label{sec:1}


Populations of self-sustained oscillators can exhibit various synchronization phenomena
\cite{ref:winfree80,ref:kuramoto84,ref:pikovsky01,ref:strogatz03,ref:manrubia04}.
For example,
it is well known that a limit-cycle oscillator can exhibit phase locking to a periodic external forcing;
this phenomenon is called the forced synchronization~\cite{ref:winfree80,ref:kuramoto84,ref:pikovsky01}.
Recently,
it was also found that uncoupled identical limit-cycle oscillators subject to weak common noise
can exhibit in-phase synchronization;
this remarkable phenomenon is called the common-noise-induced synchronization
\cite{ref:teramae04,ref:goldobin05,ref:nakao07,ref:kurebayashi12}.
In general,
each oscillatory dynamics is described by a stable limit-cycle solution to an ordinary differential equation,
and the phase description method for ordinary limit-cycle oscillators
has played an essential role in the theoretical analysis of the synchronization phenomena
\cite{ref:winfree80,ref:kuramoto84,ref:pikovsky01,
  ref:hoppensteadt97,ref:izhikevich07,ref:ermentrout10,ref:ermentrout96,ref:brown04}.
On the basis of the phase description,
optimization methods for the dynamical properties of limit-cycle oscillators have also been developed for
forced synchronization~\cite{ref:moehlis06,ref:harada10,ref:dasanayake11,ref:zlotnik12,ref:zlotnik13} and
common-noise-induced synchronization~\cite{ref:marella08,ref:abouzeid09,ref:hata11}.

Synchronization phenomena of spatiotemporal rhythms
described by partial differential equations,
such as reaction-diffusion equations and fluid equations,
have also attracted considerable attention
\cite{ref:pikovsky01,ref:manrubia04,ref:mikhailov06,ref:mikhailov13}
(see also Refs.~\cite{ref:manneville90,ref:cross93,ref:cross09} for the spatiotemporal pattern formation).
Examples of earlier studies include the following.
In reaction-diffusion systems,
synchronization between two locally coupled domains
of excitable media exhibiting spiral waves has been experimentally investigated
using the photosensitive Belousov-Zhabotinsky reaction~\cite{ref:hildebrand03}.
In fluid systems,
synchronization in both periodic and chaotic regimes has been experimentally investigated
using a periodically forced rotating fluid annulus~\cite{ref:read09}
and a pair of thermally coupled rotating fluid annuli~\cite{ref:read10}.
Of particular interest in this paper is
the experimental study on generalized synchronization of spatiotemporal chaos
in a liquid crystal spatial light modulator~\cite{ref:rogers04};
this experimental synchronization can be considered as
common-noise-induced synchronization of spatiotemporal chaos.
However, detailed theoretical analysis of these synchronization phenomena has not been performed
even for the case in which
the spatiotemporal rhythms are described by stable limit-cycle solutions to partial differential equations,
because a phase description method for partial differential equations has not been fully developed yet.

In this paper,
we theoretically analyze common-noise-induced phase synchronization
between uncoupled identical Hele-Shaw cells exhibiting oscillatory convection;
the oscillatory convection is described by a stable limit-cycle solution to a partial differential equation.
A Hele-Shaw cell is a rectangular cavity in which the gap between two vertical walls
is much smaller than the other two spatial dimensions,
and the fluid in the cavity exhibits oscillatory convection under appropriate parameter conditions
(see Refs.~\cite{ref:bernardini04,ref:nield06} and also references therein).
In Ref.~\cite{ref:kawamura13},
we recently formulated a theory for the phase description of oscillatory convection in the Hele-Shaw cell
and analyzed the mutual synchronization between a pair of coupled systems of oscillatory Hele-Shaw convection;
the theory can be considered as an extension of our phase description method
for stable limit-cycle solutions to nonlinear Fokker-Planck equations~\cite{ref:kawamura11}
(see also Ref.~\cite{ref:nakao12} for the phase description of spatiotemporal rhythms in reaction-diffusion equations).
Using the phase description method for oscillatory convection,
we here demonstrate that uncoupled systems of oscillatory Hele-Shaw convection
can be in-phase synchronized by applying weak common noise.
Furthermore, we develop a method for obtaining the optimal spatial pattern of the common noise
to achieve synchronization.
The theoretical results are validated by direct numerical simulations of the oscillatory Hele-Shaw convection.

This paper is organized as follows.
In Sec.~\ref{sec:2},
we briefly review our phase description method for oscillatory convection in the Hele-Shaw cell.
In Sec.~\ref{sec:3},
we theoretically analyze common-noise-induced phase synchronization of the oscillatory convection.
In Sec.~\ref{sec:4},
we confirm our theoretical results by numerical analysis of the oscillatory convection.
Concluding remarks are given in Sec.~\ref{sec:5}.

\section{Phase description method for oscillatory convection} \label{sec:2}

In this section, for the sake of readability and being self-contained,
we review governing equations for oscillatory convection in the Hele-Shaw cell
and our phase description method for the oscillatory convection
with consideration of its application to common-noise-induced synchronization.
More details and other applications of the phase description method are given in Ref.~\cite{ref:kawamura13}.

\subsection{Dimensionless form of the governing equations}

The dynamics of the temperature field $T(x, y, t)$ in the Hele-Shaw cell
is described by the following dimensionless form
(see Ref.~\cite{ref:bernardini04} and also references therein):
\begin{align}
  \frac{\partial}{\partial t} T(x, y, t)
  = \nabla^2 T + J(\psi, T).
  \label{eq:T}
\end{align}
The Laplacian and Jacobian are respectively given by
\begin{align}
  \nabla^2 T
  &= \left( \frac{\partial^2}{\partial x^2}
  + \frac{\partial^2}{\partial y^2} \right) T,
  \\
  J(\psi, T)
  &= \frac{\partial \psi}{\partial x} \frac{\partial T}{\partial y}
  - \frac{\partial \psi}{\partial y} \frac{\partial T}{\partial x}.
\end{align}
The stream function $\psi(x, y, t)$ is determined from the temperature field $T(x, y, t)$ as
\begin{align}
  \nabla^2 \psi(x, y, t) = -{\rm Ra} \frac{\partial T}{\partial x},
  \label{eq:P_T}
\end{align}
where the Rayleigh number is denoted by ${\rm Ra}$.
The system is defined in the unit square: $x \in [0, 1]$ and $y \in [0, 1]$.
The boundary conditions for the temperature field $T(x, y, t)$ are given by
\begin{align}
  \left. \frac{\partial T(x, y, t)}{\partial x} \right|_{x = 0}  =
  \left. \frac{\partial T(x, y, t)}{\partial x} \right|_{x = 1} &= 0,
  \label{eq:bcTx} \\
  \Bigl. T(x, y, t) \Bigr|_{y = 0}  = 1, \qquad
  \Bigl. T(x, y, t) \Bigr|_{y = 1} &= 0,
  \label{eq:bcTy}
\end{align}
where the temperature at the bottom ($y = 0$) is higher than that at the top ($y = 1$).
The stream function $\psi(x, y, t)$ satisfies
the Dirichlet zero boundary condition on both $x$ and $y$, i.e.,
\begin{align}
  \Bigl. \psi(x, y, t) \Bigr|_{x = 0}  =
  \Bigl. \psi(x, y, t) \Bigr|_{x = 1} &= 0,
  \label{eq:bcPx} \\
  \Bigl. \psi(x, y, t) \Bigr|_{y = 0}  =
  \Bigl. \psi(x, y, t) \Bigr|_{y = 1} &= 0.
  \label{eq:bcPy}
\end{align}

To simplify the boundary conditions in Eq.~(\ref{eq:bcTy}),
we consider the convective component $X(x, y, t)$ of the temperature field $T(x, y, t)$ as follows:
\begin{align}
  T(x, y, t) = (1 - y) + X(x, y, t).
  \label{eq:T_X}
\end{align}
Inserting Eq.~(\ref{eq:T_X}) into Eqs.~(\ref{eq:T})(\ref{eq:P_T}),
we derive the following equation for the convective component $X(x, y, t)$:
\begin{align}
  \frac{\partial}{\partial t} X(x, y, t)
  = \nabla^2 X + J(\psi, X) - \frac{\partial \psi}{\partial x},
  \label{eq:X}
\end{align}
where the stream function $\psi(x, y, t)$ is determined by
\begin{align}
  \nabla^2 \psi(x, y, t) = -{\rm Ra} \frac{\partial X}{\partial x}.
  \label{eq:P_X}
\end{align}
Applying Eq.~(\ref{eq:T_X}) to Eqs.~(\ref{eq:bcTx})(\ref{eq:bcTy}),
we obtain the following boundary conditions for the convective component $X(x, y, t)$:
\begin{align}
  \left. \frac{\partial X(x, y, t)}{\partial x} \right|_{x = 0}  =
  \left. \frac{\partial X(x, y, t)}{\partial x} \right|_{x = 1} &= 0,
  \label{eq:bcXx} \\
  \Bigl. X(x, y, t) \Bigr|_{y = 0}  =
  \Bigl. X(x, y, t) \Bigr|_{y = 1} &= 0.
  \label{eq:bcXy}
\end{align}
That is, the convective component $X(x, y, t)$ satisfies
the Neumann zero boundary condition on $x$
and the Dirichlet zero boundary condition on $y$.
It should be noted that
this system does not possess translational or rotational symmetry owing to the boundary conditions
given by Eqs.~(\ref{eq:bcPx})(\ref{eq:bcPy})(\ref{eq:bcXx})(\ref{eq:bcXy}).

\subsection{Limit-cycle solution and its Floquet zero eigenfunctions}

The dependence of the Hele-Shaw convection on the Rayleigh number ${\rm Ra}$ is well known,
and the existence of stable limit-cycle solutions to Eq.~(\ref{eq:X}) is also well established
(see Ref.~\cite{ref:bernardini04} and also references therein).
In general, a stable limit-cycle solution to Eq.~(\ref{eq:X}),
which represents oscillatory convection in the Hele-Shaw cell,
can be described by
\begin{align}
  X(x, y, t) = X_0\bigl( x, y, \Theta(t) \bigr), \qquad
  \dot{\Theta}(t) = \Omega.
  \label{eq:X_X0}
\end{align}
The phase and natural frequency are denoted by $\Theta$ and $\Omega$, respectively.
The limit-cycle solution $X_0(x, y, \Theta)$
possesses the following $2\pi$-periodicity in $\Theta$:
$X_0(x, y, \Theta + 2\pi) = X_0(x, y, \Theta)$.
Inserting Eq.~(\ref{eq:X_X0}) into Eqs.~(\ref{eq:X})(\ref{eq:P_X}),
we find that the limit-cycle solution $X_0(x, y, \Theta)$ satisfies
\begin{align}
  \Omega \frac{\partial}{\partial \Theta} X_0(x, y, \Theta)
  = \nabla^2 X_0 + J(\psi_0, X_0) - \frac{\partial \psi_0}{\partial x},
  \label{eq:X0}
\end{align}
where the stream function $\psi_0(x, y, \Theta)$ is determined by
\begin{align}
  \nabla^2 \psi_0(x, y, \Theta) = -{\rm Ra} \frac{\partial X_0}{\partial x}.
  \label{eq:P0}
\end{align}
From Eq.~(\ref{eq:T_X}), the corresponding temperature field $T_0(x, y, \Theta)$ is given by
(e.g., see Fig.~\ref{fig:2} in Sec.~\ref{sec:4})
\begin{align}
  T_0(x, y, \Theta) = (1 - y) + X_0(x, y, \Theta).
  \label{eq:T0}
\end{align}

Let $u(x, y, \Theta, t)$ represent a small disturbance added to the limit-cycle solution $X_0(x, y, \Theta)$,
and consider a slightly perturbed solution
\begin{align}
  X(x, y, t) = X_0\bigl( x, y, \Theta(t) \bigr) + u\bigl( x, y, \Theta(t), t \bigr).
\end{align}
Equation~(\ref{eq:X}) is then linearized with respect to $u(x, y, \Theta, t)$ as follows:
\begin{align}
  \frac{\partial}{\partial t} u(x, y, \Theta, t)
  = {\cal L}(x, y, \Theta) u(x, y, \Theta, t).
  \label{eq:linear}
\end{align}
As in the limit-cycle solution $X_0(x, y, \Theta)$,
the function $u(x, y, \Theta)$ satisfies
the Neumann zero boundary condition on $x$
and the Dirichlet zero boundary condition on $y$.
Note that ${\cal L}(x, y, \Theta)$ is time-periodic through $\Theta$.
Therefore, Eq.~(\ref{eq:linear}) is a Floquet-type system with a periodic linear operator.
Defining the inner product of two functions as
\begin{align}
  \Pd{ u^\ast(x, y, \Theta), \, u(x, y, \Theta) }
  = \frac{1}{2\pi} \int_0^{2\pi} d\Theta \int_0^1 dx \int_0^1 dy \,
  u^\ast(x, y, \Theta) u(x, y, \Theta),
  \label{eq:inner}
\end{align}
we introduce the adjoint operator of the linear operator ${\cal L}(x, y, \Theta)$ by
\begin{align}
  \Pd{ u^\ast(x, y, \Theta), \, {\cal L}(x, y, \Theta) u(x, y, \Theta) }
  = \Pd{ {\cal L}^\ast(x, y, \Theta) u^\ast(x, y, \Theta), \, u(x, y, \Theta) }.
  \label{eq:operator}
\end{align}
As in $u(x, y, \Theta)$,
the function $u^\ast(x, y, \Theta)$ also satisfies
the Neumann zero boundary condition on $x$
and the Dirichlet zero boundary condition on $y$.
Details of the derivation of the adjoint operator ${\cal L}^\ast(x, y, \Theta)$
are given in Ref.~\cite{ref:kawamura13}.

In the following subsection,
we utilize the Floquet eigenfunctions associated with the zero eigenvalue, i.e.,
\begin{align}
  {\cal L}(x, y, \Theta) U_0(x, y, \Theta)
  &= 0,
  \\
  {\cal L}^\ast(x, y, \Theta) U_0^\ast(x, y, \Theta)
  &= 0.
\end{align}
We note that the right zero eigenfunction $U_0(x, y, \Theta)$ can be chosen as
\begin{align}
  U_0(x, y, \Theta) = \frac{\partial}{\partial \Theta} X_0(x, y, \Theta),
  \label{eq:U0}
\end{align}
which is confirmed by differentiating Eq.~(\ref{eq:X0}) with respect to $\Theta$.
Using the inner product of Eq.~(\ref{eq:inner})
with the right zero eigenfunction of Eq.~(\ref{eq:U0}),
the left zero eigenfunction $U_0^\ast(x, y, \Theta)$ is normalized as
\begin{align}
  \Pd{ U_0^\ast(x, y, \Theta), \, U_0(x, y, \Theta) }
  = \frac{1}{2\pi} \int_0^{2\pi} d\Theta \int_0^1 dx \int_0^1 dy \,
  U_0^\ast(x, y, \Theta) U_0(x, y, \Theta)
  = 1.
\end{align}
Here, we can show that the following equation holds
(see also Refs.~\cite{ref:hoppensteadt97,ref:kawamura13,ref:kawamura11}):
\begin{align}
  \frac{\partial}{\partial \Theta}
  \left[ \int_0^1 dx \int_0^1 dy \, U_0^\ast(x, y, \Theta) U_0(x, y, \Theta) \right] = 0.
\end{align}
Therefore, the following normalization condition
is satisfied independently for each $\Theta$ as follows:
\begin{equation}
  \int_0^1 dx \int_0^1 dy \, U_0^\ast(x, y, \Theta) U_0(x, y, \Theta) = 1.
\end{equation}

\subsection{Oscillatory convection under weak perturbations}

We now consider oscillatory Hele-Shaw convection
with a weak perturbation applied to the temperature field $T(x, y, t)$
described by the following equation:
\begin{align}
  \frac{\partial}{\partial t} T(x, y, t)
  = \nabla^2 T + J(\psi, T) + \epsilon p(x, y, t).
  \label{eq:T_p}
\end{align}
The weak perturbation is denoted by $\epsilon p(x, y, t)$.
Inserting Eq.~(\ref{eq:T_X}) into Eq.~(\ref{eq:T_p}),
we obtain the following equation for the convective component $X(x, y, t)$:
\begin{align}
  \frac{\partial}{\partial t} X(x, y, t)
  = \nabla^2 X + J(\psi, X) - \frac{\partial \psi}{\partial x} + \epsilon p(x, y, t).
  \label{eq:X_p}
\end{align}
Using the idea of the phase reduction~\cite{ref:kuramoto84},
we can derive a phase equation from the perturbed equation~(\ref{eq:X_p}).
Namely, we project the dynamics of the perturbed equation~(\ref{eq:X_p})
onto the unperturbed solution as
\begin{align}
  \dot{\Theta}(t)
  &= \int_0^1 dx \int_0^1 dy \, U_0^\ast(x, y, \Theta)
  \left[ \frac{\partial}{\partial t} X(x, y, t) \right]
  \nonumber \\
  &= \int_0^1 dx \int_0^1 dy \, U_0^\ast(x, y, \Theta)
  \left[ \nabla^2 X + J(\psi, X) - \frac{\partial \psi}{\partial x} + \epsilon p(x, y, t) \right]
  \nonumber \\
  &\simeq \int_0^1 dx \int_0^1 dy \, U_0^\ast(x, y, \Theta)
  \left[ \nabla^2 X_0 + J(\psi_0, X_0) - \frac{\partial \psi_0}{\partial x} + \epsilon p(x, y, t) \right]
  \nonumber \\
  &= \int_0^1 dx \int_0^1 dy \, U_0^\ast(x, y, \Theta)
  \left[ \Omega \frac{\partial}{\partial \Theta} X_0(x, y, \Theta) + \epsilon p(x, y, t) \right]
  \nonumber \\
  &= \int_0^1 dx \int_0^1 dy \, U_0^\ast(x, y, \Theta)
  \, \biggl[ \Omega \, U_0(x, y, \Theta) + \epsilon p(x, y, t) \biggr]
  \nonumber \\
  &= \Omega + \epsilon \int_0^1 dx \int_0^1 dy \, U_0^\ast(x, y, \Theta) p(x, y, t),
\end{align}
where we approximated $X(x, y, t)$ by the unperturbed limit-cycle solution $X_0(x, y, \Theta)$.
Therefore, the phase equation describing the oscillatory Hele-Shaw convection with a weak perturbation
is approximately obtained in the following form:
\begin{align}
  \dot{\Theta}(t) = \Omega + \epsilon \int_0^1 dx \int_0^1 dy \, Z(x, y, \Theta) p(x, y, t),
  \label{eq:Theta_p}
\end{align}
where the {\it phase sensitivity function} is defined as
(e.g., see Fig.~\ref{fig:2} in Sec.~\ref{sec:4})
\begin{equation}
  Z(x, y, \Theta) = U_0^\ast(x, y, \Theta).
\end{equation}
Here, we note that
the phase sensitivity function $Z(x, y, \Theta)$ satisfies
the Neumann zero boundary condition on $x$
and the Dirichlet zero boundary condition on $y$, i.e.,
\begin{align}
  \left. \frac{\partial Z(x, y, \Theta)}{\partial x} \right|_{x = 0}  =
  \left. \frac{\partial Z(x, y, \Theta)}{\partial x} \right|_{x = 1} &= 0,
  \label{eq:bcZx} \\
  \Bigl. Z(x, y, \Theta) \Bigr|_{y = 0}  =
  \Bigl. Z(x, y, \Theta) \Bigr|_{y = 1} &= 0.
  \label{eq:bcZy}
\end{align}
As mentioned in Ref.~\cite{ref:kawamura13},
Eq.~(\ref{eq:Theta_p}) is a generalization of the phase equation for a perturbed limit-cycle oscillator
described by a finite-dimensional dynamical system
(see Refs.~\cite{ref:winfree80,ref:kuramoto84,ref:pikovsky01,
  ref:hoppensteadt97,ref:izhikevich07,ref:ermentrout10,ref:ermentrout96,ref:brown04}).
However, reflecting the aspects of an infinite-dimensional dynamical system,
the phase sensitivity function $Z(x, y, \Theta)$ of the oscillatory Hele-Shaw convection
possesses infinitely many components that are continuously parameterized by the two variables, $x$ and $y$.

In this paper,
we further consider the case that the perturbation is described by a product of two functions as follows:
\begin{align}
  p(x, y, t) = a(x, y) q(t).
  \label{eq:p_aq}
\end{align}
That is, the space-dependence and time-dependence of the perturbation are separated.
In this case, the phase equation~(\ref{eq:Theta_p}) can be written in the following form:
\begin{align}
  \dot{\Theta}(t) = \Omega + \epsilon \zeta(\Theta) q(t),
  \label{eq:Theta_q}
\end{align}
where the {\it effective phase sensitivity function} is given by
(e.g., see Fig.~\ref{fig:5} in Sec.~\ref{sec:4})
\begin{align}
  \zeta(\Theta) = \int_0^1 dx \int_0^1 dy \, Z(x, y, \Theta) a(x, y).
  \label{eq:zeta}
\end{align}
We note that the form of Eq.~(\ref{eq:Theta_q}) is essentially the same as
that of the phase equation for a perturbed limit-cycle oscillator
described by a finite-dimensional dynamical system
(see Refs.~\cite{ref:winfree80,ref:kuramoto84,ref:pikovsky01,
  ref:hoppensteadt97,ref:izhikevich07,ref:ermentrout10,ref:ermentrout96,ref:brown04}).
We also note that the effective phase sensitivity function $\zeta(\Theta)$
can also be considered as the collective phase sensitivity function
in the context of the collective phase description
of coupled individual dynamical elements exhibiting macroscopic rhythms~\cite{ref:kawamura11,ref:kawamura08,ref:kawamura10}.

\section{Theoretical analysis of the common-noise-induced synchronization} \label{sec:3}

In this section,
using the phase description method in Sec.~\ref{sec:2},
we analytically investigate common-noise-induced synchronization
between uncoupled systems of oscillatory Hele-Shaw convection.
In particular,
we theoretically determine the optimal spatial pattern of the common noise for achieving the noise-induced synchronization.

\subsection{Phase reduction and Lyapunov exponent}

We consider $N$ uncoupled systems of oscillatory Hele-Shaw convection subject to weak common noise
described by the following equation for $\sigma = 1, \cdots, N$:
\begin{align}
  \frac{\partial}{\partial t} T_\sigma(x, y, t)
  = \nabla^2 T_\sigma + J(\psi_\sigma, T_\sigma)
  + \epsilon a(x, y) \xi(t),
  \label{eq:T_xi}
\end{align}
where the weak common noise is denoted by $\epsilon a(x, y) \xi(t)$.
Inserting Eq.~(\ref{eq:T_X}) into Eq.~(\ref{eq:T_xi}) for each $\sigma$,
we obtain the following equation for the convective component $X_\sigma(x, y, t)$:
\begin{align}
  \frac{\partial}{\partial t} X_\sigma(x, y, t)
  = \nabla^2 X_\sigma + J(\psi_\sigma, X_\sigma) - \frac{\partial \psi_\sigma}{\partial x}
  + \epsilon a(x, y) \xi(t).
  \label{eq:X_xi}
\end{align}
As in Eq.~(\ref{eq:P_X}), the stream function of each system is determined by
\begin{align}
  \nabla^2 \psi_\sigma(x, y, t)
  = -{\rm Ra} \frac{\partial X_\sigma}{\partial x}.
\end{align}
The common noise $\xi(t)$
is assumed to be white Gaussian noise~\cite{ref:risken89,ref:gardiner97},
the statistics of which are given by
\begin{align}
  \langle \xi(t) \rangle = 0, \qquad
  \langle \xi(t) \xi(s) \rangle = 2\delta(t - s).
  \label{eq:xi}
\end{align}
Here, we assume that the unperturbed oscillatory Hele-Shaw convection is a stable limit cycle
and that the noise intensity $\epsilon^2$ is sufficiently weak.
Then, as in Eq.~(\ref{eq:Theta_q}),
we can derive a phase equation from Eq.~(\ref{eq:X_xi}) as follows~\footnote{
  Precisely speaking, owing to the noise,
  the frequency of the oscillatory convection given in Eq.~(\ref{eq:Theta_xi})
  can be slightly different from the natural frequency given in Eq.~(\ref{eq:X_X0});
  however, this point is not essential in this paper
  because Eq.~(\ref{eq:Lambda}) is independent of the value of the frequency.
  The theory of stochastic phase reduction for ordinary limit-cycle oscillators has been intensively investigated
  in Refs.~\cite{ref:yoshimura08,ref:teramae09,ref:nakao10,ref:goldobin10},
  but extensions to partial differential equations have not been developed yet.
}:
\begin{align}
  \dot{\Theta}_\sigma(t) = \Omega + \epsilon \zeta(\Theta_\sigma) \xi(t),
  \label{eq:Theta_xi}
\end{align}
where the effective phase sensitivity function $\zeta(\Theta)$ is given by Eq.~(\ref{eq:zeta}).
Once the phase equation~(\ref{eq:Theta_xi}) is obtained,
the Lyapunov exponent characterizing the common-noise-induced synchronization
can be derived using the argument by Teramae and Tanaka~\cite{ref:teramae04}.
From Eqs.~(\ref{eq:xi})(\ref{eq:Theta_xi}),
the Lyapunov exponent,
which quantifies the exponential growth rate of small phase differences between the two systems,
can be written in the following form:
\begin{align}
  \Lambda = -\frac{\epsilon^2}{2\pi} \int_0^{2\pi} d\Theta \, \Bigl[ \zeta'(\Theta) \Bigr]^2 \leq 0.
  \label{eq:Lambda}
\end{align}
Here, we used the following abbreviation: $\zeta'(\Theta) = d\zeta(\Theta)/d\Theta$.
Equation~(\ref{eq:Lambda}) represents that
uncoupled systems of oscillatory Hele-Shaw convection
can be in-phase synchronized when driven by the weak common noise,
as long as the phase reduction approximation is valid.
In the following two subsections,
we develop a method for obtaining the optimal spatial pattern of the common noise
to achieve the noise-induced synchronization of the oscillatory convection.

\subsection{Spectral decomposition of the phase sensitivity function}

Considering the boundary conditions of $Z(x, y, \Theta)$, Eqs.~(\ref{eq:bcZx})(\ref{eq:bcZy}),
we introduce the following spectral transformation~\footnote{
  Practically speaking, e.g., in numerical simulations,
  infinite series are truncated at some sufficiently large finite number.
  From a theoretical point of view,
  such a truncation approximation is valid
  because this system includes dissipation due to the Laplacian.
}:
\begin{align}
  Z_{jk}(\Theta) = \int_0^1 dx \int_0^1 dy \, Z(x, y, \Theta) \cos(\pi j x) \sin(\pi k y),
\end{align}
for $j = 0, 1, 2, \cdots$ and $k = 1, 2, \cdots$.
The corresponding spectral decomposition of $Z(x, y, \Theta)$ is given by
\begin{align}
  Z(x, y, \Theta) = 4 \sum_{j=0}^\infty \sum_{k=1}^\infty Z_{jk}(\Theta) \cos(\pi j x) \sin(\pi k y).
  \label{eq:Z_CosSin}
\end{align}
By inserting Eq.~(\ref{eq:Z_CosSin}) into Eq.~(\ref{eq:zeta}),
the effective phase sensitivity function $\zeta(\Theta)$ can be written in the following form:
\begin{align}
  \zeta(\Theta)
  = \int_0^1 dx \int_0^1 dy \, Z(x, y, \Theta) a(x, y)
  = \sum_{j=0}^\infty \sum_{k=1}^\infty b_{jk} Z_{jk}(\Theta),
  \label{eq:zeta_double}
\end{align}
where the spectral transformation of $a(x, y)$ is defined as
\begin{align}
  b_{jk} = 4 \int_0^1 dx \int_0^1 dy \, a(x, y) \cos(\pi j x) \sin(\pi k y).
\end{align}
The corresponding spectral decomposition of $a(x, y)$ is given by
\begin{align}
  a(x, y) = \sum_{j=0}^\infty \sum_{k=1}^\infty b_{jk} \cos(\pi j x) \sin(\pi k y).
\end{align}
For the sake of convenience in the calculation below,
we rewrite the double sum in Eq.~(\ref{eq:zeta_double})
by the following single series:
\begin{align}
  \zeta(\Theta)
  = \sum_{j=0}^\infty \sum_{k=1}^\infty b_{jk} Z_{jk}(\Theta)
  \equiv \sum_{n=0}^\infty s_n Q_n(\Theta).
  \label{eq:zeta_single}
\end{align}
In Eq.~(\ref{eq:zeta_single}),
we introduced one-dimensional representations,
$s_n = b_{jk}$ and $Q_n(\Theta) = Z_{jk}(\Theta)$,
where the mapping between $n$ and $(j, k)$ is bijective.
Accordingly, we obtain the following quantity:
\begin{align}
  \Bigl[ \zeta'(\Theta) \Bigr]^2
  = \sum_{n=0}^\infty \sum_{m=0}^\infty s_n s_m Q_n'(\Theta) Q_m'(\Theta),
  \label{eq:dzeta_squared}
\end{align}
where $Q_n'(\Theta) = dQ_n(\Theta)/d\Theta$.
From Eqs.~(\ref{eq:Lambda})(\ref{eq:dzeta_squared}),
the Lyapunov exponent normalized by the noise intensity, $-\Lambda / \epsilon^2$,
can be written in the following form:
\begin{align}
  -\frac{\Lambda}{\epsilon^2}
  = \frac{1}{2\pi} \int_0^{2\pi} d\Theta \, \Bigl[ \zeta'(\Theta) \Bigr]^2
  = \sum_{n=0}^\infty \sum_{m=0}^\infty K_{nm} s_n s_m,
  \label{eq:Lambda_K}
\end{align}
where each element of the symmetric matrix $\hat{K}$ is given by
\begin{align}
  K_{nm} = \frac{1}{2\pi} \int_0^{2\pi} d\Theta \, Q_n'(\Theta) Q_m'(\Theta) = K_{mn}.
  \label{eq:K}
\end{align}

\subsection{Spectral components of the optimal spatial pattern} \label{subsec:3C}

By defining an infinite-dimensional column vector $\bd{s} \equiv ( s_0, s_1, s_2, \cdots )^{\rm T}$,
Eq.~(\ref{eq:Lambda_K}) can also be written as
\begin{align}
  -\frac{\Lambda}{\epsilon^2}
  = \sum_{n=0}^\infty \sum_{m=0}^\infty K_{nm} s_n s_m
  = \bd{s} \cdot \hat{K} \bd{s},
  \label{eq:Lyapunov}
\end{align}
which is a quadratic form.
Using the spectral representation of the normalized Lyapunov exponent, Eq.~(\ref{eq:Lyapunov}),
we seek the optimal spatial pattern of the common noise for the synchronization.
As a constraint, we introduce the following condition:
\begin{align}
  \bd{s} \cdot \bd{s}
  = \sum_{n=0}^\infty s_n^2
  = \sum_{j=0}^\infty \sum_{k=1}^\infty b_{jk}^2
  = 1.
  \label{eq:unity}
\end{align}
That is, the total power of the spatial pattern is fixed at unity.
Under this constraint condition, we consider the maximization of Eq.~(\ref{eq:Lyapunov}).
For this purpose, we define the Lagrangian $F(\bd{s}, \lambda)$ as
\begin{align}
  F(\bd{s}, \lambda)
  = \sum_{n=0}^\infty \sum_{m=0}^\infty K_{nm} s_n s_m - \lambda \left( \sum_{n=0}^\infty s_n^2 - 1 \right),
\end{align}
where the Lagrange multiplier is denoted by $\lambda$.
Setting the derivative of the Lagrangian $F(\bd{s}, \lambda)$ to be zero,
we can obtain the following equations:
\begin{align}
  \frac{\partial F}{\partial s_l}
  &= 2 \left( \sum_{m=0}^\infty K_{l m} s_m - \lambda s_l \right)
  = 0, \qquad (\, l = 0, 1, 2, \cdots \,),
  \\[1mm]
  \frac{\partial F}{\partial \lambda}
  &= - \left( \sum_{n=0}^\infty s_n^2 - 1 \right)
  = 0,
\end{align}
which are equivalent to the eigenvalue problem described by
\begin{align}
  \hat{K} \bd{s}_{\alpha} = \lambda_\alpha \bd{s}_\alpha, \qquad
  \bd{s}_\alpha \cdot \bd{s}_\alpha = 1, \qquad
  (\, \alpha = 0, 1, 2, \cdots \,).
\end{align}
These eigenvectors $\bd{s}_\alpha$ and the corresponding eigenvalues $\lambda_\alpha$ satisfy
\begin{align}
  F(\bd{s}_\alpha, \lambda_\alpha) = \lambda_\alpha.
\end{align}
Because the matrix $\hat{K}$, which is defined in Eq.~(\ref{eq:K}), is symmetric,
the eigenvalues $\lambda_\alpha$ are real numbers.
Consequently, under the constraint condition given by Eq.~(\ref{eq:unity}),
the optimal vector that maximizes Eq.~(\ref{eq:Lambda})
coincides with the eigenvector associated with the largest eigenvalue, i.e.,
\begin{align}
  \lambda_{\rm opt} = \max_{\alpha} \, \lambda_\alpha.
\end{align}
Therefore, the optimal spatial pattern $a_{\rm opt}(x, y)$ can be written in the following form:
\begin{align}
  a_{\rm opt}(x, y) = \sum_{j=0}^\infty \sum_{k=1}^\infty b_{\rm opt}(j, k) \cos(\pi j x) \sin(\pi k y),
  \label{eq:aopt}
\end{align}
where the coefficients $b_{\rm opt}(j, k)$ in the double series
correspond to the elements of the optimal vector $\bd{s}_{\rm opt}$ associated with $\lambda_{\rm opt}$.
From Eq.~(\ref{eq:Lyapunov}), the Lyapunov exponent is then given by
\begin{align}
  \Lambda_{\rm opt} = - \epsilon^2 \lambda_{\rm opt}.
\end{align}
Finally, we note that this optimization method can also be considered as
the principal component analysis~\cite{ref:jolliffe02}
of the phase-derivative of the phase sensitivity function, $\partial_\Theta Z(x, y, \Theta)$.

\section{Numerical analysis of the common-noise-induced synchronization} \label{sec:4}

In this section,
to illustrate the theory developed in Sec.~\ref{sec:3},
we numerically investigate common-noise-induced synchronization
between uncoupled Hele-Shaw cells exhibiting oscillatory convection.
The numerical simulation method is summarized in Ref.~\footnote{
  We applied the pseudospectral method,
  which is composed of
  a sine expansion with $128$ modes for the Dirichlet zero boundary condition
  and a cosine expansion with $128$ modes for the Neumann zero boundary condition.
  The fourth-order Runge-Kutta method with integrating factor
  using a time step $\varDelta t = 10^{-4} \,\sim\, 10^{-6}$ (mainly, $\varDelta t = 10^{-4}$)
  and the Heun method with integrating factor
  using a time step $\varDelta t = 10^{-5}$
  were applied for the deterministic and stochastic (Langevin-type) equations, respectively.
}.

\subsection{Spectral decomposition of the convective component}

Considering the boundary conditions of the convective component $X(x, y, \Theta)$, Eqs.~(\ref{eq:bcXx})(\ref{eq:bcXy}),
we introduce the following spectral transformation:
\begin{align}
  H_{jk}(t) = \int_0^1 dx \int_0^1 dy \, X(x, y, t) \cos(\pi j x) \sin(\pi k y),
\end{align}
for $j = 0, 1, 2, \cdots$ and $k = 1, 2, \cdots$.
The corresponding spectral decomposition of the convective component $X(x, y, \Theta)$ is given by
\begin{align}
  X(x, y, t) = 4 \sum_{j=0}^\infty \sum_{k=1}^\infty H_{jk}(t) \cos(\pi j x) \sin(\pi k y).
\end{align}
In visualizing the limit-cycle orbit in the infinite-dimensional state space,
we project the limit-cycle solution $X_0(x, y, \Theta)$ onto the $H_{11}$-$H_{22}$ plane as
\begin{align}
  H_{11}(\Theta)
  &= \int_0^1 dx \int_0^1 dy \, X_0(x, y, \Theta) \cos(\pi x) \sin(\pi y),
  \\
  H_{22}(\Theta)
  &= \int_0^1 dx \int_0^1 dy \, X_0(x, y, \Theta) \cos(2 \pi x) \sin(2 \pi y).
\end{align}

\subsection{Limit-cycle solution and phase sensitivity function}

\begin{figure*}
  \begin{center}
    \includegraphics[width=0.5\hsize,clip]{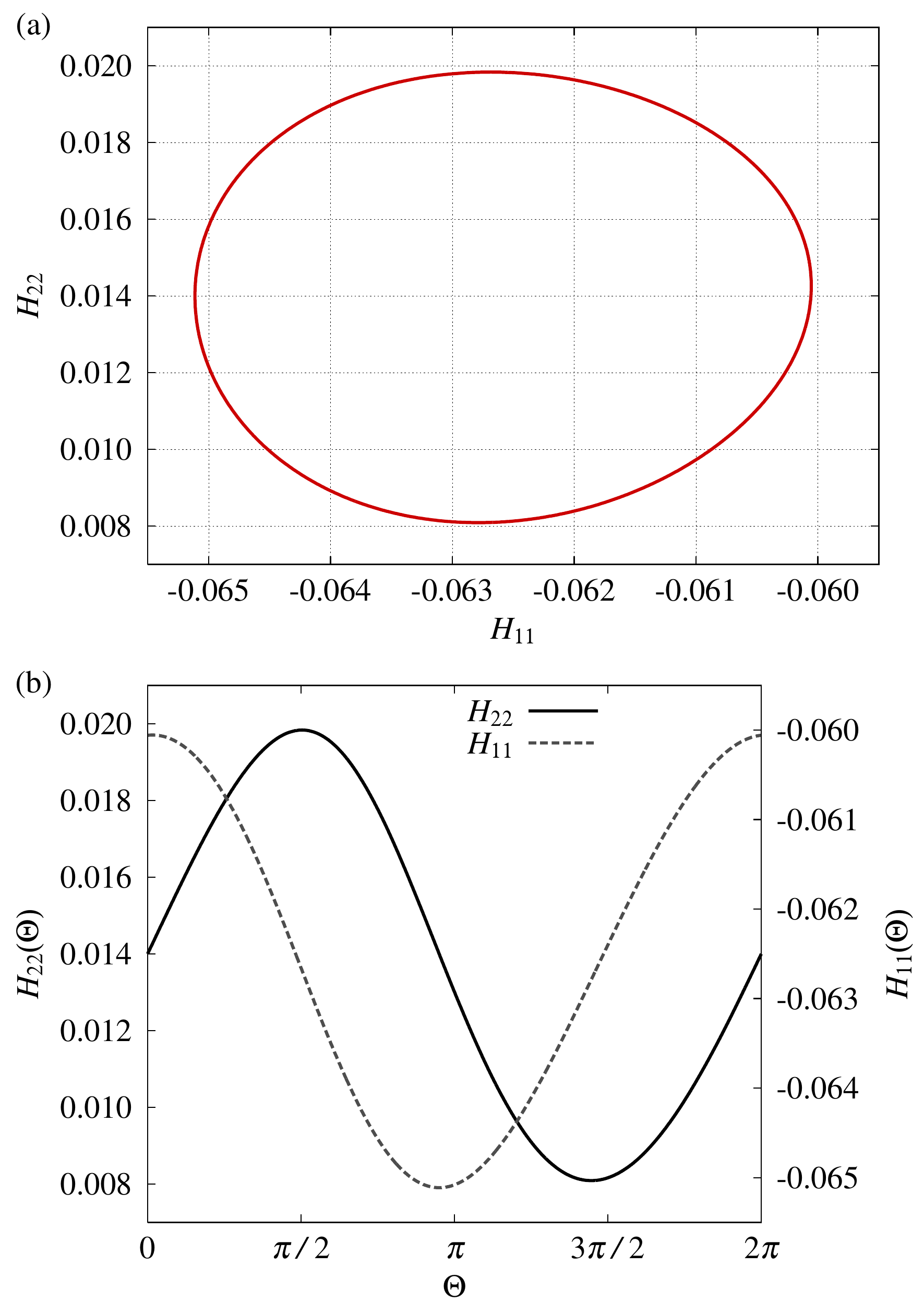}
  \end{center}
  \caption{(Color online)
    (a) Limit-cycle orbit projected onto the $H_{11}$-$H_{22}$ plane.
    (b) Waveforms of $H_{11}(\Theta)$ and $H_{22}(\Theta)$.
    The Rayleigh number is ${\rm Ra} = 480$,
    and then the natural frequency is $\Omega \simeq 622$,
    i.e., the oscillation period is $2\pi / \Omega \simeq 0.010$.
  }
  \label{fig:1}
\end{figure*}

\begin{figure*}
  \begin{center}
    \includegraphics[width=\hsize,clip]{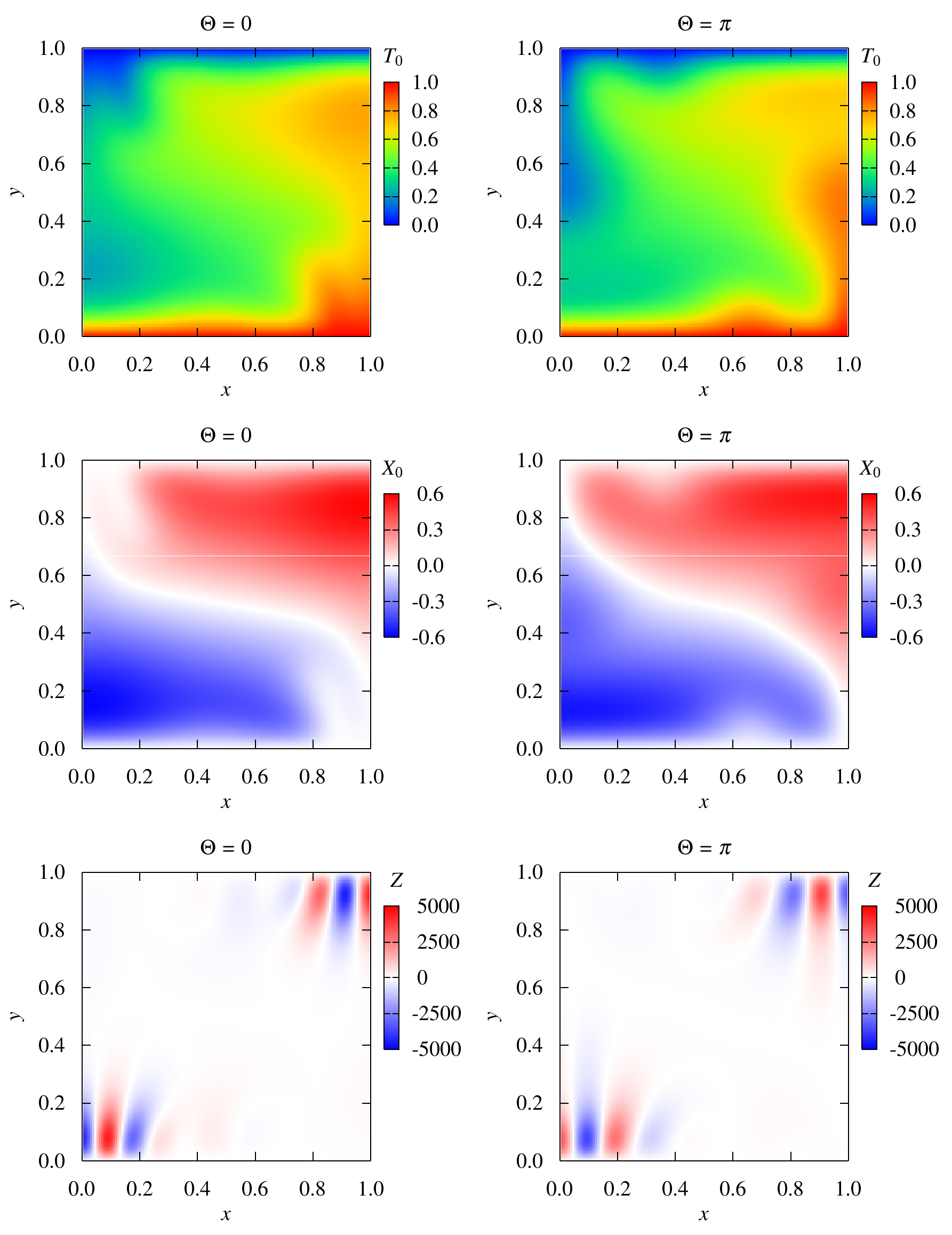}
  \end{center}
  \caption{(Color online)
    Snapshots of
    $T_0(x, y, \Theta)$,
    $X_0(x, y, \Theta)$, and
    $Z(x, y, \Theta)$ for
    $\Theta = 0$ and $\Theta = \pi$.
  }
  \label{fig:2}
\end{figure*}

The initial values were prepared
so that the system exhibits single cellular oscillatory convection.
The Rayleigh number was fixed at ${\rm Ra} = 480$,
which gives the natural frequency $\Omega \simeq 622$,
i.e., the oscillation period $2\pi/\Omega \simeq 0.010$.
Figure~\ref{fig:1} shows the limit-cycle orbit of the oscillatory convection
projected onto the $H_{11}$-$H_{22}$ plane,
obtained from direct numerical simulations of the dynamical equation~(\ref{eq:X}).
Snapshots of the limit-cycle solution $X_0(x, y, \Theta)$ and other associated functions,
$T_0(x, y, \Theta)$ and $Z(x, y, \Theta)$, are shown in Fig.~\ref{fig:2},
where the phase variable $\Theta$ is discretized using $512$ grid points.
We note that Fig.~\ref{fig:1} and Fig.~\ref{fig:2} are essentially
reproductions of our previous results given in Ref.~\cite{ref:kawamura13}.
Details of the numerical method for obtaining the phase sensitivity function $Z(x, y, \Theta)$
are given in Refs.~\cite{ref:kawamura13,ref:kawamura11}
(see also Refs.~\cite{ref:hoppensteadt97,ref:izhikevich07,ref:ermentrout10,ref:ermentrout96,ref:brown04}).

As seen in Fig.~\ref{fig:2},
the phase sensitivity function $Z(x, y, \Theta)$ is spatially localized.
Namely, the absolute values of the phase sensitivity function $Z(x, y, \Theta)$
in the top-right and bottom-left corner regions of the system
are much larger than those in the other regions;
this fact reflects the dynamics of the spatial pattern of the convective component $X_0(x, y, \Theta)$.

As mentioned in Ref.~\cite{ref:kawamura13},
the phase sensitivity function $Z(x, y, \Theta)$ in this case possesses the following symmetry.
For each $\Theta$,
the limit-cycle solution $X_0(x, y, \Theta)$ and the phase sensitivity function $Z(x, y, \Theta)$,
shown in Fig.~\ref{fig:2},
are anti-symmetric with respect to the center of the system, i.e.,
\begin{align}
  X_0(-x_\delta, -y_\delta, \Theta)
  &= -X_0(x_\delta, y_\delta, \Theta),
  \\
  Z(-x_\delta, -y_\delta, \Theta)
  &= -Z(x_\delta, y_\delta, \Theta),
  \label{eq:anti-symmetric}
\end{align}
where $x_\delta = x - 1/2$ and $y_\delta = y - 1/2$.
Therefore, for a spatial pattern $a_{\rm s}(x, y)$
that is symmetric with respect to the center of the system,
\begin{align}
  a_{\rm s}(-x_\delta, -y_\delta)
  = a_{\rm s}(x_\delta, y_\delta),
\end{align}
the corresponding effective phase sensitivity function $\zeta(\Theta)$ becomes zero, i.e.,
\begin{align}
  \zeta(\Theta)
  = \int_0^1 dx \int_0^1 dy \, Z(x, y, \Theta) a_{\rm s}(x, y)
  = 0.
  \label{eq:zeta_s}
\end{align}
That is, such symmetric perturbations do not affect the phase of the oscillatory convection.

\subsection{Optimal spatial pattern of the common noise}

The optimal spatial pattern is obtained as the best combination of single-mode spatial patterns,
i.e., Eq.~(\ref{eq:aopt}).
Thus, we first consider the following single-mode spatial pattern:
\begin{align}
  a(x, y) = a_{(j, k)}(x, y) \equiv \cos(\pi j x) \sin(\pi k y).
\end{align}
Then, the effective phase sensitivity function is given by the following single spectral component:
\begin{align}
  \zeta(\Theta)
  = \int_0^1 dx \int_0^1 dy \, Z(x, y, \Theta) \cos(\pi j x) \sin(\pi k y)
  = Z_{jk}(\Theta).
  \label{eq:Zjk}
\end{align}
From Eq.~(\ref{eq:Lambda}),
the Lyapunov exponent for the single-mode spatial pattern can be written in the following form:
\begin{align}
  \Lambda(j, k) = -\frac{\epsilon^2}{2\pi} \int_0^{2\pi} d\Theta \, \Bigl[ Z_{jk}'(\Theta) \Bigr]^2,
\end{align}
where $Z_{jk}'(\Theta) = dZ_{jk}(\Theta)/d\Theta$.

\begin{figure*}
  \begin{center}
    \includegraphics[width=\hsize,clip]{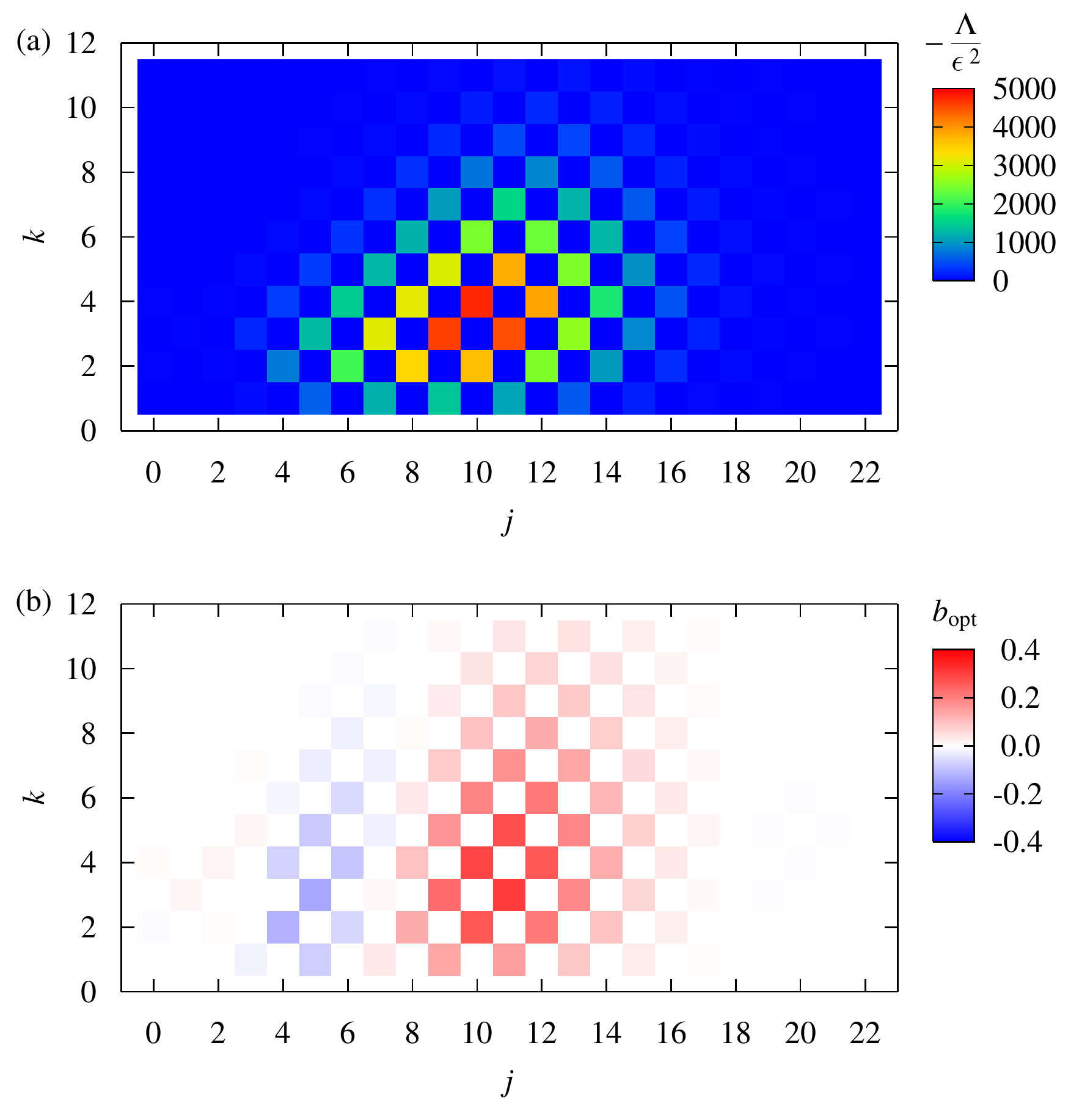}
  \end{center}
  \caption{(Color online)
    (a) Normalized Lyapunov exponent for single-mode spatial patterns, $-\Lambda(j, k) / \epsilon^2$,
    i.e., spatial power spectrum of $\partial_\Theta Z(x, y, \Theta)$ averaged over $\Theta$.
    (b) Spectral components of the optimal spatial pattern, i.e., $b_{\rm opt}(j, k)$.
  }
  \label{fig:3}
\end{figure*}

\begin{figure*}
  \begin{center}
    \includegraphics[width=\hsize,clip]{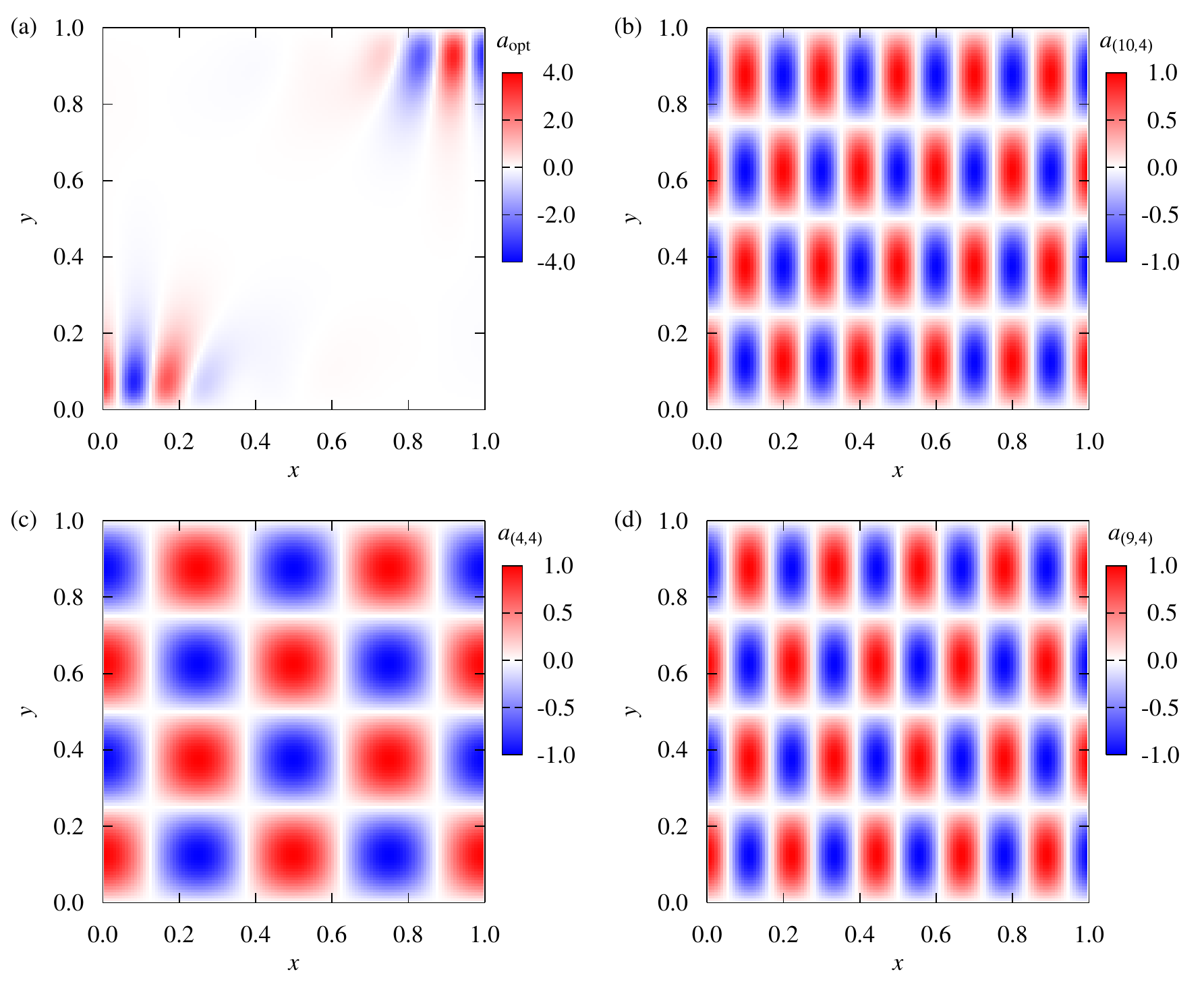}
  \end{center}
  \caption{(Color online)
    (a) Optimal spatial pattern
    $a_{\rm opt}(x, y) = \sum_{j=0}^\infty \sum_{k=1}^\infty b_{\rm opt}(j, k) \cos(\pi j x) \sin(\pi k y)$.
    (b) Single-mode spatial pattern $a_{(10, 4)}(x, y) = \cos(10 \pi x) \sin(4 \pi y)$.
    (c) Single-mode spatial pattern $a_{( 4, 4)}(x, y) = \cos( 4 \pi x) \sin(4 \pi y)$.
    (d) Single-mode spatial pattern $a_{( 9, 4)}(x, y) = \cos( 9 \pi x) \sin(4 \pi y)$.
  }
  \label{fig:4}
\end{figure*}

Figure~\ref{fig:3}(a) shows the normalized Lyapunov exponent for single-mode spatial patterns,
i.e., $-\Lambda(j, k) / \epsilon^2$.
Owing to the anti-symmetry of the phase sensitivity function, given in Eq.~(\ref{eq:anti-symmetric}),
the normalized Lyapunov exponent $-\Lambda(j, k) / \epsilon^2$ exhibits a checkerboard pattern,
namely, $-\Lambda(j, k)/ \epsilon^2 = 0$ when the sum of $j$ and $k$, i.e., $j + k$, is an odd number.
The maximum of $-\Lambda(j, k) / \epsilon^2$ is located at $(j, k) = (10, 4)$;
under the condition of $j = k$,
the maximum of $-\Lambda(j, k) / \epsilon^2$ is located at $(j, k) = ( 4, 4)$.
The single-mode spatial patterns,
$a_{(10, 4)}(x, y)$,
$a_{( 4, 4)}(x, y)$, and
$a_{( 9, 4)}(x, y)$,
are shown in Figs.~\ref{fig:4}(b)(c)(d), respectively.
We note that
$a_{(10, 4)}(x, y)$ and $a_{( 4, 4)}(x, y)$ are anti-symmetric with respect to the center of the system,
whereas $a_{( 9, 4)}(x, y)$ is symmetric.
These spatial patterns are used in the numerical simulations performed below.

We now consider the optimal spatial pattern.
Figure~\ref{fig:3}(b) shows the spectral components of the optimal spatial pattern, i.e., $b_{\rm opt}(j, k)$,
obtained by the optimization method developed in Sec.~\ref{subsec:3C};
Figure~\ref{fig:4}(a) shows the corresponding optimal spatial pattern, i.e., $a_{\rm opt}(x, y)$,
given by Eq.~(\ref{eq:aopt}).
As seen in Fig.~\ref{fig:3},
when the normalized Lyapunov exponent for a single-mode spatial pattern, $-\Lambda(j, k) / \epsilon^2$, is large,
the absolute value of the optimal spectral components, $|b_{\rm opt}(j, k)|$, is also large.
As seen in Fig.~\ref{fig:4}(a),
the optimal spatial pattern $a_{\rm opt}(x, y)$ is similar to
the snapshots of the phase sensitivity function $Z(x, y, \Theta)$ shown in Fig.~\ref{fig:2}.
In fact, as mentioned in Sec.~\ref{subsec:3C},
the optimal spatial pattern $a_{\rm opt}(x, y)$ corresponds to
the first principal component of $\partial_\Theta Z(x, y, \Theta)$.
Reflecting the anti-symmetry of the phase sensitivity function, Eq.~(\ref{eq:anti-symmetric}),
the optimal spatial pattern $a_{\rm opt}(x, y)$ is also anti-symmetric with respect to the center of the system.

\subsection{Effective phase sensitivity function}

\begin{figure*}
  \begin{center}
    \includegraphics[width=\hsize,clip]{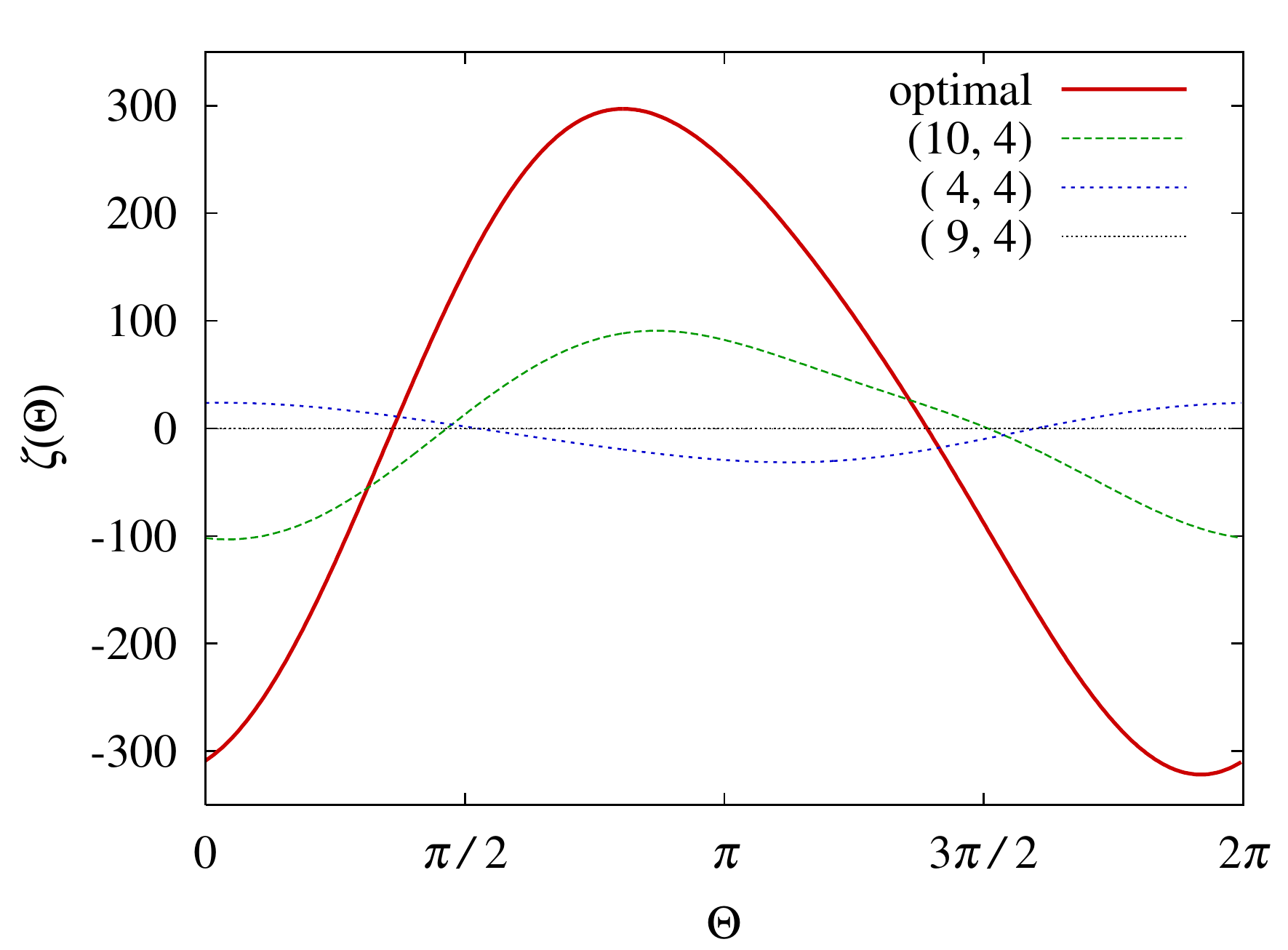}
  \end{center}
  \caption{(Color online)
    Effective phase sensitivity function $\zeta(\Theta)$ for the following spatial patterns:
    $a_{\rm opt}(x, y)$,
    $a_{(10, 4)}(x, y)$,
    $a_{( 4, 4)}(x, y)$, and
    $a_{( 9, 4)}(x, y)$,
    which are shown in Fig.~\ref{fig:4}.
  }
  \label{fig:5}
\end{figure*}

Figure~\ref{fig:5} shows the effective phase sensitivity functions $\zeta(\Theta)$
for the spatial patterns shown in Fig.~\ref{fig:4}.
When the normalized Lyapunov exponent $-\Lambda(j, k) / \epsilon^2$ is large,
the amplitude of the corresponding effective phase sensitivity function $\zeta(\Theta)$ is also large.
For the spatial pattern $a_{( 9, 4)}(x, y)$,
which is symmetric with respect to the center of the system,
the effective phase sensitivity function becomes zero, $\zeta(\Theta) = 0$,
as shown in Eq.~(\ref{eq:zeta_s}).

To confirm the theoretical results shown in Fig.~\ref{fig:5},
we obtain the effective phase sensitivity function $\zeta(\Theta)$
by direct numerical simulations of Eq.~(\ref{eq:X_p}) with Eq.~(\ref{eq:p_aq}) as follows:
we measure the phase response of the oscillatory convection
by applying a weak impulsive perturbation with the spatial pattern $a(x, y)$
to the limit-cycle solution $X_0(x, y, \Theta)$ with the phase $\Theta$;
then, normalizing the phase response curve by the weak impulse intensity $\epsilon$,
we obtain the effective phase sensitivity function $\zeta(\Theta)$.
The effective phase sensitivity function $\zeta(\Theta)$
obtained by direct numerical simulations with impulse intensity $\epsilon$
are compared with the theoretical curves in Fig.~\ref{fig:6}.
The simulation results agree quantitatively with the theory~\footnote{
  When the impulsive perturbation is sufficiently weak,
  the phase response curve depends linearly on the impulse intensity $\epsilon$.
  Therefore, the phase response curve normalized by the impulse intensity $\epsilon$
  converges to the effective phase sensitivity function $\zeta(\Theta)$ as $\epsilon$ decreases.
  As shown in Fig.~\ref{fig:6}(d), when the impulsive perturbation is not weak,
  the dependence of the phase response curve on the impulse intensity $\epsilon$ becomes nonlinear.
  In general, when the impulsive perturbation is not weak, the phase response curve is not equal to zero,
  even though the effective phase sensitivity function is equal to zero, $\zeta(\Theta) = 0$.
  We also note that the linear dependence region of the phase response curve on the impulse
  is generally dependent on the spatial pattern $a(x, y)$ of the impulse.
}.

\subsection{Common-noise-induced synchronization}

In this subsection,
we demonstrate the common-noise-induced synchronization
between uncoupled Hele-Shaw cells exhibiting oscillatory convection
by direct numerical simulations of the stochastic (Langevin-type) partial differential equation~(\ref{eq:X_xi}).
Theoretical values of
both the Lyapunov exponents $\Lambda$ for several spatial patterns $a(x, y)$
with the common noise intensity $\epsilon^2 = 10^{-6}$
and the corresponding relaxation time $1 / |\Lambda|$ toward the synchronized state
are summarized in Table~\ref{table:1}.

\begin{figure*}
  \begin{center}
    \includegraphics[width=\hsize,clip]{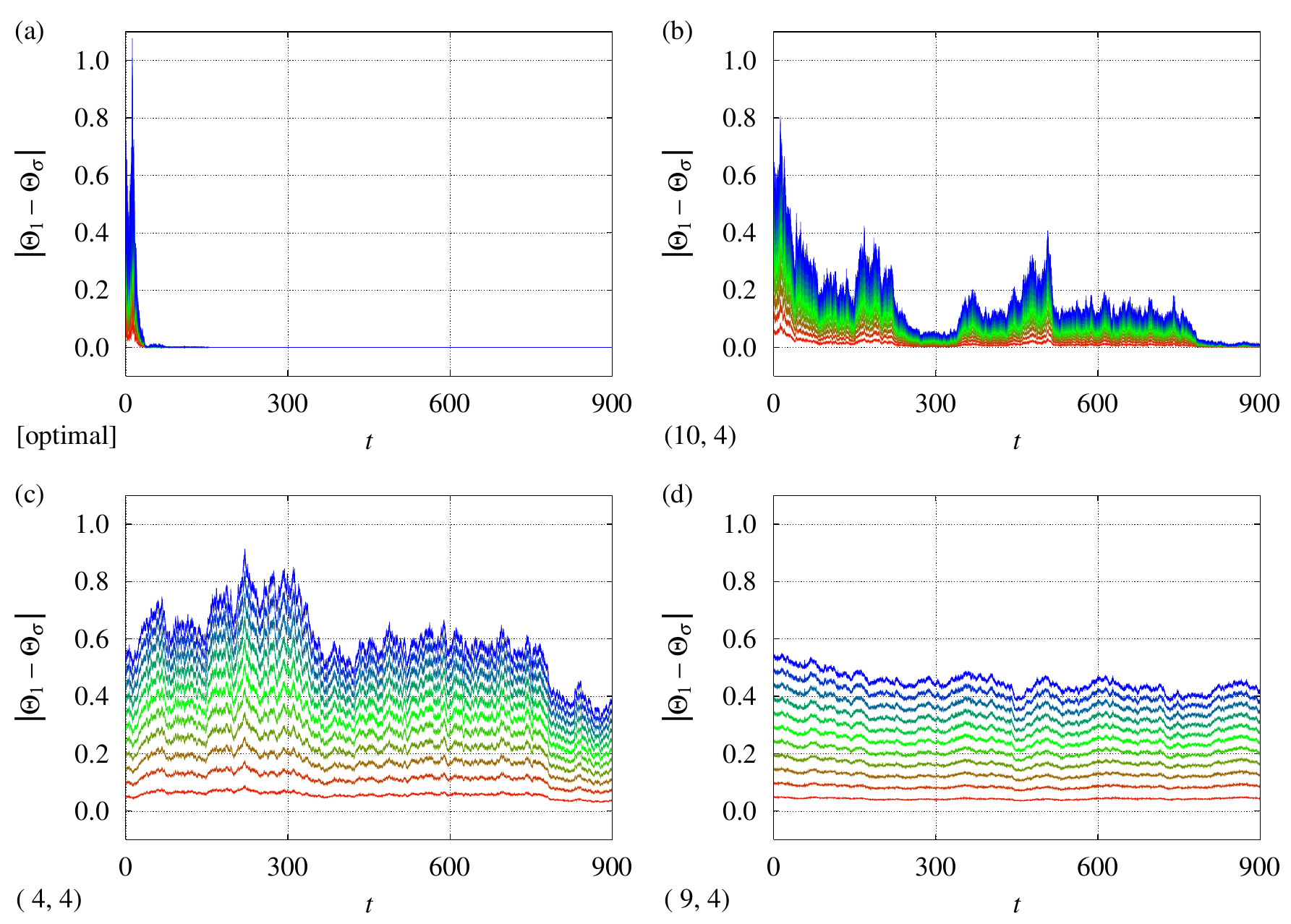}
  \end{center}
  \caption{(Color online)
    Time evolution of phase differences $| \Theta_1 - \Theta_\sigma |$
    with the common noise intensity $\epsilon^2 = 10^{-6}$.
    The initial phases are $\Theta_\sigma(t = 0) = 2 \pi (\sigma - 1) / 128$
    for $\sigma = 1, \cdots, 12$.
    The spatial patterns of common noise are as follows.
    (a) $a_{\rm opt}(x, y)$.
    (b) $a_{(10, 4)}(x, y)$.
    (c) $a_{( 4, 4)}(x, y)$.
    (d) $a_{( 9, 4)}(x, y)$.
  }
  \label{fig:7}
\end{figure*}

\begin{figure*}
  \begin{center}
    \includegraphics[width=\hsize,clip]{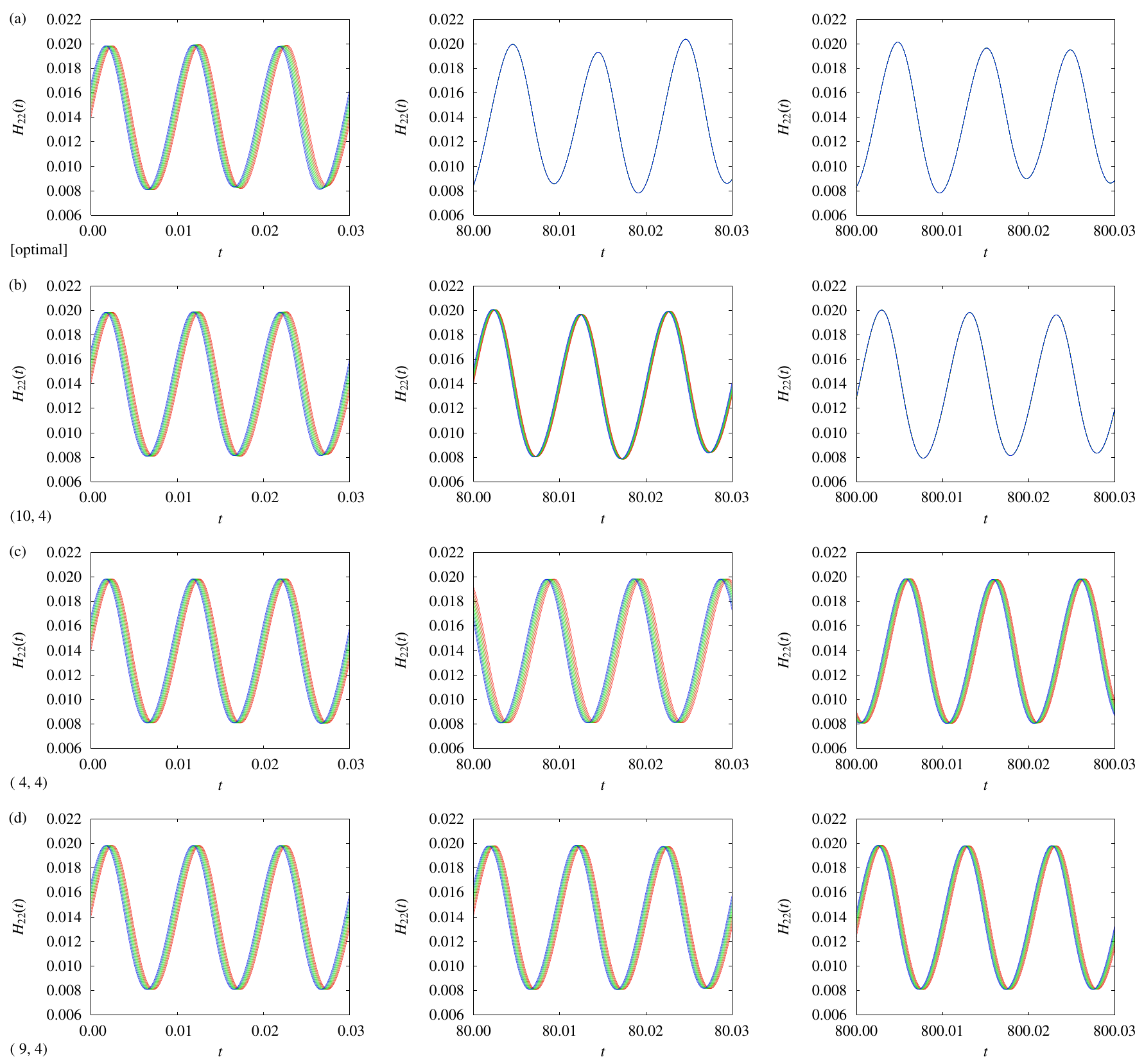}
  \end{center}
  \caption{(Color online)
    Time evolution of $H_{22}^{(\sigma)}(t)$,
    which corresponds to Fig.~\ref{fig:7},
    for the following spatial patterns.
    (a) $a_{\rm opt}(x, y)$.
    (b) $a_{(10, 4)}(x, y)$.
    (c) $a_{( 4, 4)}(x, y)$.
    (d) $a_{( 9, 4)}(x, y)$.
  }
  \label{fig:8}
\end{figure*}

Figure~\ref{fig:7} shows
the time evolution of the phase differences $| \Theta_1 - \Theta_\sigma |$
when the common noise intensity is $\epsilon^2 = 10^{-6}$.
The initial phase values are $\Theta_\sigma(t = 0) = 2 \pi (\sigma - 1) / 128$
for $\sigma = 1, \cdots, 12$.
Figure~\ref{fig:8} shows
the time evolution of $H_{22}^{(\sigma)}(t)$,
which corresponds to Fig.~\ref{fig:7}.
The relaxation times estimated from the simulation results agree reasonably well with the theory~\footnote{
  Theoretically speaking,
  the phase differences shown in Fig.~\ref{fig:7}(d) should be constant
  because the effective phase sensitivity function is equal to zero, $\zeta(\Theta) = 0$, for this case.
  As shown in Fig.~\ref{fig:6}(d),
  when the perturbation is not sufficiently weak, the phase response curve is not equal to zero;
  this higher order effect causes the slight variations shown in Fig.~\ref{fig:7}(d).
}.
As seen in Fig.~\ref{fig:7} and Fig.~\ref{fig:8},
the relaxation time for the optimal spatial pattern $a_{\rm opt}(x, y)$
is actually much smaller than those for the single-mode spatial patterns.
For the cases of single-mode patterns,
the relaxation time for the single-mode spatial pattern $a_{(10, 4)}(x, y)$
is also smaller than those for the other single-mode spatial patterns,
$a_{( 4, 4)}(x, y)$ and $a_{( 9, 4)}(x, y)$.
We also note that
the time evolution of both $| \Theta_1 - \Theta_\sigma |$ and $H_{22}^{(\sigma)}(t)$ for $a_{(10, 4)}(x, y)$
is significantly different from that for $a_{( 9, 4)}(x, y)$
in spite of the similarity between the two spatial patterns of the neighboring modes;
this difference results from the difference of symmetry with respect to the center,
as shown in Eq.~(\ref{eq:zeta_s}).

\begin{figure*}
  \begin{center}
    \includegraphics[width=\hsize,clip]{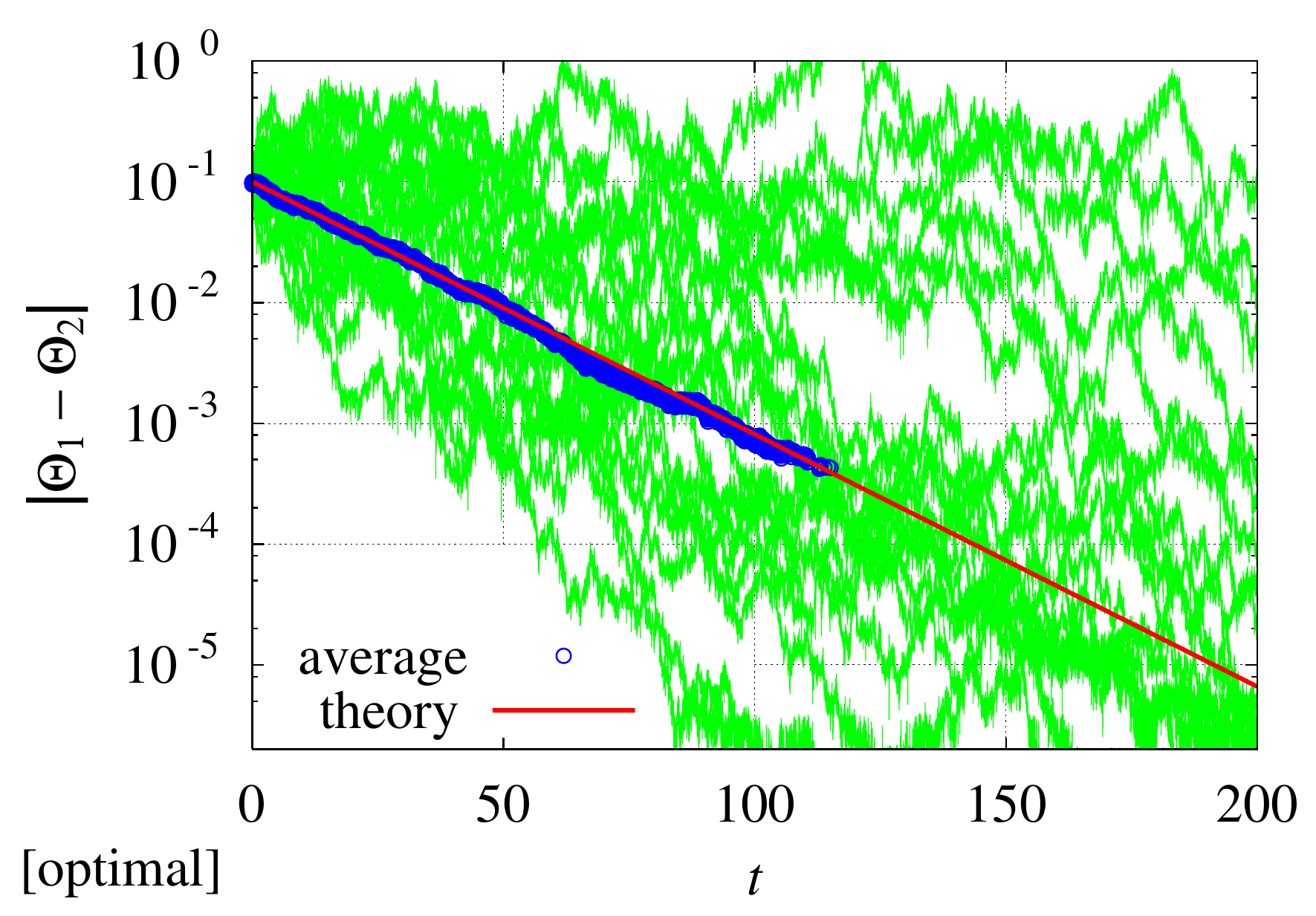}
  \end{center}
  \caption{(Color online)
    Comparison of the Lyapunov exponent for the optimal spatial pattern $a_{\rm opt}(x, y)$
    with the common noise intensity $\epsilon^2 = 10^{-6}$
    between direct numerical simulations (average) and the theoretical curve (theory).
    The results of direct numerical simulations are averaged over $100$ samples,
    in which only $20$ samples are shown by thin (green) lines.
    The averaged result is shown only up to the time
    when one of the phase differences in these $100$ samples numerically converges to zero.
    The initial phases are $\Theta_\sigma(t = 0) = 2 \pi (\sigma - 1) / 64$ for $\sigma = 1, 2$,
    i.e., the initial phase difference is $| \Theta_1(t = 0) - \Theta_2(t = 0) | \simeq 10^{-1}$.
  }
  \label{fig:9}
\end{figure*}

Figure~\ref{fig:9} shows a quantitative comparison of the Lyapunov exponents
between direct numerical simulations and the theory
for the case of the optimal spatial pattern $a_{\rm opt}(x, y)$.
The initial phase values are $\Theta_\sigma(t = 0) = 2 \pi (\sigma - 1) / 64$ for $\sigma = 1, 2$,
i.e., the initial phase difference is $| \Theta_1(t = 0) - \Theta_2(t = 0) | \simeq 10^{-1}$.
The results of direct numerical simulations are averaged over $100$ samples for different noise realizations.
The simulation results quantitatively agree with the theory.

\begin{figure*}
  \begin{center}
    \includegraphics[width=\hsize,clip]{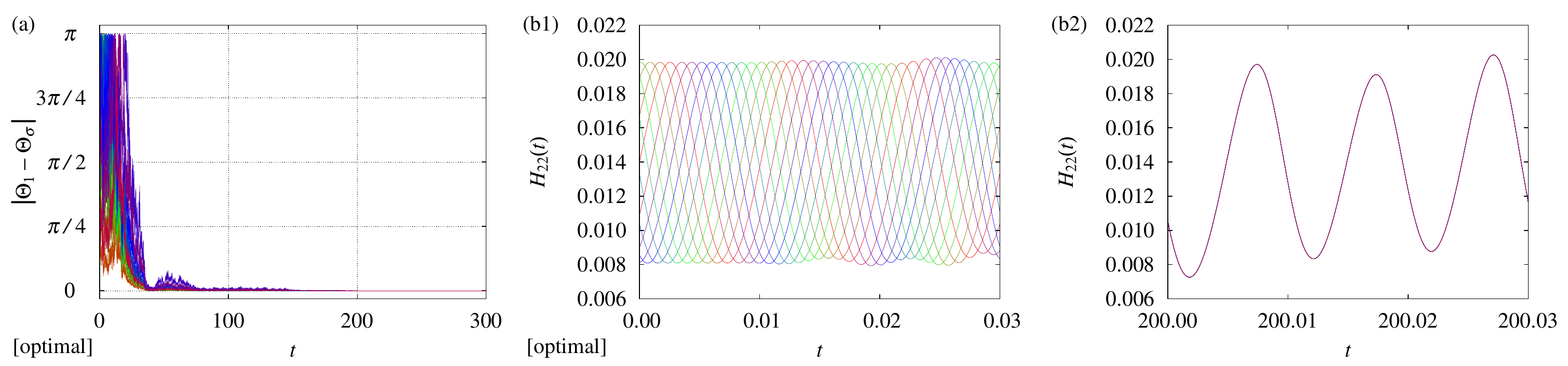}
  \end{center}
  \caption{(Color online)
    Global stability for the optimal spatial pattern $a_{\rm opt}(x, y)$
    with the common noise intensity $\epsilon^2 = 10^{-6}$.
    The initial phases are $\Theta_\sigma(t = 0) = 2 \pi (\sigma - 1) / 12$
    for $\sigma = 1, \cdots, 12$.
    (a) Time evolution of phase differences $| \Theta_1 - \Theta_\sigma |$.
    (b1)(b2) Time evolution of $H_{22}^{(\sigma)}(t)$.
  }
  \label{fig:10}
\end{figure*}

Figure~\ref{fig:10} shows
the global stability of the common-noise-induced synchronization of oscillatory convection
for the case of the optimal spatial pattern $a_{\rm opt}(x, y)$;
namely, it shows that the synchronization is eventually achieved from arbitrary initial phase differences,
i.e., $| \Theta_1(t = 0) - \Theta_\sigma(t = 0) | \in [0, \pi]$.
Although the Lyapunov exponent $\Lambda$ based on the linearization of Eq.~(\ref{eq:Theta_xi})
quantifies only the local stability of a small phase difference,
as long as the phase reduction approximation is valid,
this global stability holds true
for any spatial pattern $a(x, y)$ with a non-zero Lyapunov exponent,
namely, the Lyapunov exponent is negative, $\Lambda < 0$, as found from Eq.~(\ref{eq:Lambda}).
The global stability can be proved by the theory developed in Ref.~\cite{ref:nakao07},
i.e., by analyzing the Fokker-Planck equation equivalent to the Langevin-type phase equation~(\ref{eq:Theta_xi});
in addition, the effect of the independent noise can also be included.

\section{Concluding remarks} \label{sec:5}

Our investigations in this paper are summarized as follows.
In Sec.~\ref{sec:2},
we briefly reviewed our phase description method for oscillatory convection in the Hele-Shaw cell
with consideration of its application to common-noise-induced synchronization.
In Sec.~\ref{sec:3},
we analytically investigated common-noise-induced synchronization of oscillatory convection
using the phase description method.
In particular, we theoretically determined the optimal spatial pattern of the common noise
for the oscillatory Hele-Shaw convection.
In Sec.~\ref{sec:4},
we numerically investigated common-noise-induced synchronization of oscillatory convection;
the direct numerical simulation successfully confirmed the theoretical predictions.

\begin{figure*}
  \begin{center}
    \includegraphics[width=\hsize,clip]{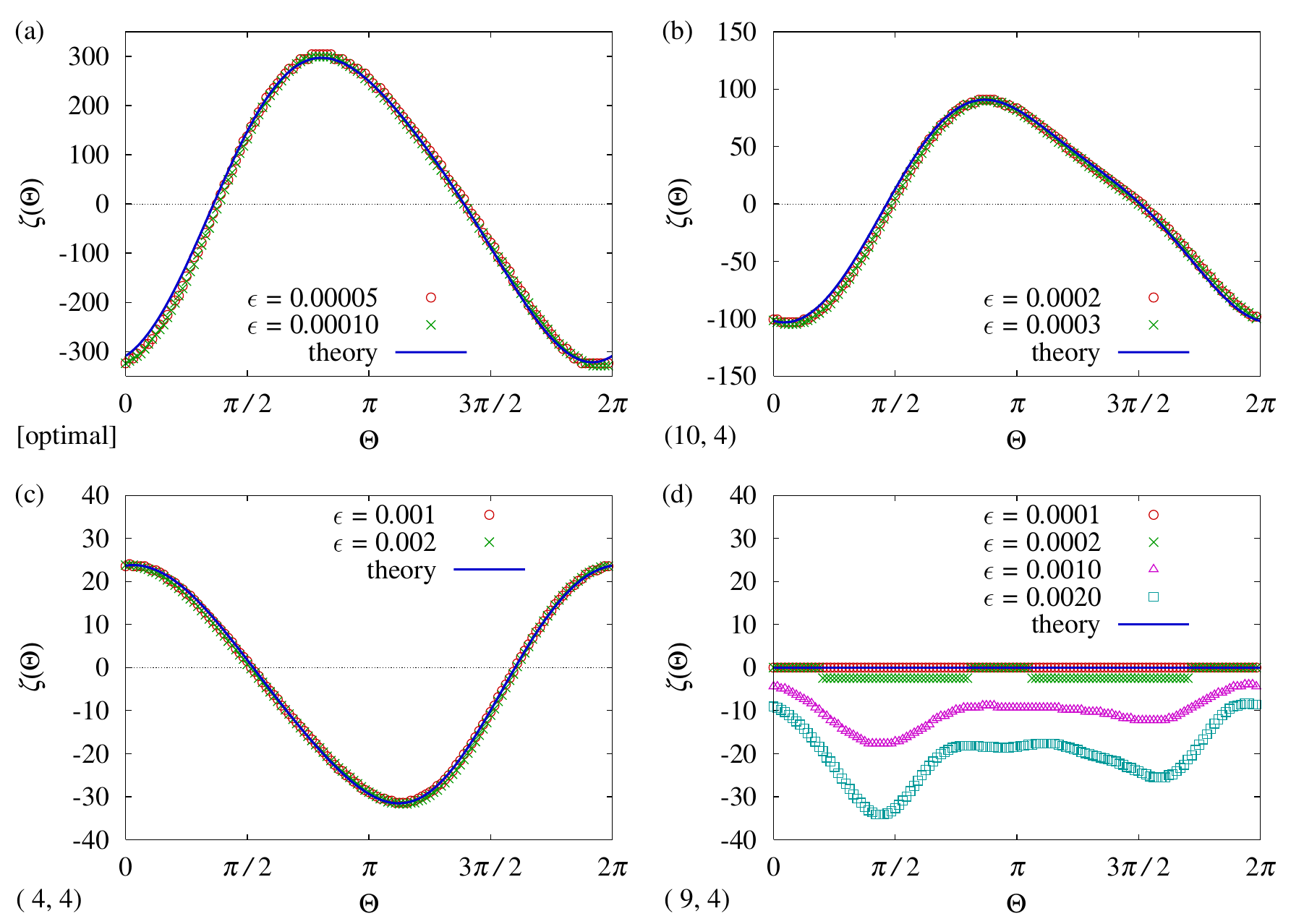}
  \end{center}
  \caption{(Color online)
    Comparisons of the effective phase sensitivity function $\zeta(\Theta)$
    between direct numerical simulations with impulse intensity $\epsilon$
    and the theoretical curve (theory) for the following spatial patterns.
    (a) $a_{\rm opt}(x, y)$.
    (b) $a_{(10, 4)}(x, y)$.
    (c) $a_{( 4, 4)}(x, y)$.
    (d) $a_{( 9, 4)}(x, y)$.
  }
  \label{fig:6}
\end{figure*}

The key quantity of the theory developed in this paper is the phase sensitivity function $Z(x, y, \Theta)$.
Thus, we describe an experimental procedure to obtain the phase sensitivity function $Z(x, y, \Theta)$.
As in Eq.~(\ref{eq:Z_CosSin}),
the phase sensitivity function $Z(x, y, \Theta)$
can be decomposed into the spectral components $Z_{jk}(\Theta)$,
which are the effective phase sensitivity functions
for the single-mode spatial patterns $a_{(j,k)}(x, y)$ as shown in Eq.~(\ref{eq:Zjk}).
In a manner similar to the direct numerical simulations yielding Fig.~\ref{fig:6},
the effective phase sensitivity function $Z_{jk}(\Theta)$
for each single-mode spatial pattern $a_{(j,k)}(x, y)$ can also be experimentally measured.
Therefore, in general, the phase sensitivity function $Z(x, y, \Theta)$
can be constructed from a sufficiently large set of such $Z_{jk}(\Theta)$.
Once the phase sensitivity function $Z(x, y, \Theta)$ is obtained,
the optimization method for common-noise-induced synchronization can also be applied in experiments.

Finally, we remark that not only the phase description method for spatiotemporal rhythms
but also the optimization method for common-noise-induced synchronization have broad applicability;
these methods are not restricted to the oscillatory Hele-Shaw convection analyzed in this paper.
For example, the combination of these methods can be applied to
common-noise-induced phase synchronization of spatiotemporal rhythms
in reaction-diffusion systems of excitable and/or heterogeneous media.
Furthermore, as mentioned above, also in experimental systems, such as
the photosensitive Belousov-Zhabotinsky reaction~\cite{ref:hildebrand03} and
the liquid crystal spatial light modulator~\cite{ref:rogers04},
the optimization method for common-noise-induced synchronization could be applied.

\begin{acknowledgments}
  Y.K. is grateful to members of both
  the Earth Evolution Modeling Research Team and
  the Nonlinear Dynamics and Its Application Research Team
  at IFREE/JAMSTEC for fruitful comments.
  Y.K. is also grateful for financial support by
  JSPS KAKENHI Grant Number 25800222.
  H.N. is grateful for financial support by
  JSPS KAKENHI Grant Numbers 25540108 and 22684020,
  CREST Kokubu project of JST, and
  FIRST Aihara project of JSPS.
\end{acknowledgments}


\clearpage


\clearpage

\begin{table*}[b]
  \caption{Lyapunov exponents for spatial patterns with the common noise intensity $\epsilon^2 = 10^{-6}$.}
  \begin{center}
    \includegraphics[width=\hsize,clip]{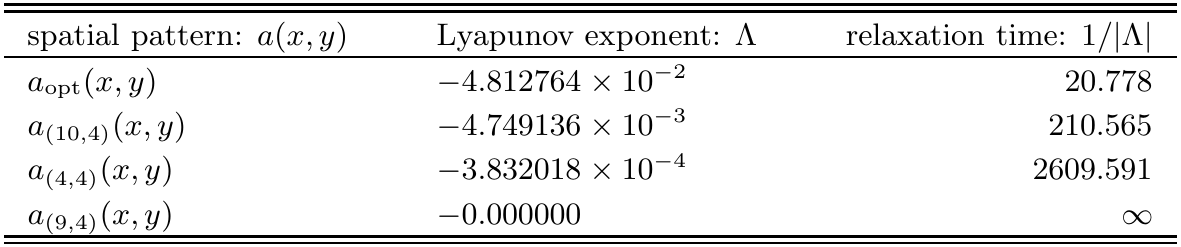}
  \end{center}
  \label{table:1}
\end{table*}

\end{document}